# Shape Changes of Self-Assembled Actin Bilayer Composite Membranes

W.Häckl, M.Bärmann and E. Sackmann

Physik Department E22 (Biophysics Laboratory), Technische Universität München, James-Franck-Straße, 85748 Garching, Germany

**We report the self-assembly of thin actin shells beneath the membranes of giant vesicles. Ion-carrier mediated influx of $Mg^{2+}$ induces actin polymerization in the initially spherical vesicles. Buckling of the vesicles and the formation of blisters after thermally induced bilayer expansion is demonstrated. Bilayer flickering is dominated by tension generated by its coupling to the actin cortex. Quantitative flicker analysis suggests the bilayer and the actin cortex are separated by 0.4 µm to 0.5 µm due to undulation forces.**

The viscoelastic properties of plasma membranes of eukaryotic cells are determined by the coupling of two soft shells: the lipid-protein bilayer and the membrane associated cytoskeleton. The latter may consist of a two-dimensional meshwork as in the case of red blood cells or of a several hundred nm thick shell (called actin cortex) composed mainly of actin and associated manipulation proteins (including crosslinking-, severing- and motor proteins) as in leucocytes or as in the amoeba such as the cells of the slime mold *Dictyostelium discoideum* (cf. [1, 2, 3]). Many biological processes are associated with cellular shape changes, such as cell locomotion or phagocytosis, and are controlled by a complex interplay of the two soft shells of the composite cell envelope. The elastic moduli of the actin cortex may be varied in two ways: (i) by the degree of crosslinking within the network and (ii) by its coupling strength to the lipid-protein bilayer. Moreover it can be actively deformed by motor proteins.

In the present work we report the reconstitution of thin actin shells in giant vesicles by self-assembly during polymerization of monomeric actin through (ion-carrier mediated) $Mg^{2+}$-influx into spherical vesicles. The actin cortex is visualized by microfluorescence using laser scanning confocal microscopy (LSCM). Buckling-like shape changes [4] are observed by phase contrast microscopy (PCM) after thermal expansion of the bilayer excess area.

Giant vesicles were prepared by swelling in an electric field [5,6] of thin lipid layers (DMPC [7] and 5 mole % of the ion carrier A23187 [8]) deposited onto ITO glass electrodes [9] in the presence of monomeric actin (G-actin) [10] and in non-polymerizing buffer ($G^x$-buffer [11]). For the LSCM experiments, vesicles were swollen in $G^x$-buffer containing 63 mOsm glucose and 4 µM rhodaminyl phalloidin [12]. The vesicles were diluted into $G^x$-buffer containing isoosmolar sucrose in order to reduce the extravesicular G-actin below the critical concentration of polymerization [10]. Polymerization of the actin inside the vesicles was then started by addition of 2 mM of $Mg^{2+}$. For some experiments the actin outside the vesicles was inactivated by adding isomolar DNase I [8]. During polymerization the vesicles were kept spherical by temperature controlled adjustment of the area-to-volume ratio. For observation the vesicles were transferred into a temperature controlled observation chamber in which they could be observed either by PCM (Zeiss Axiomat 10) or by LSCM (Axiovert 100 TV, Zeiss, Oberkochen, Germany and Noran Odyssey XL, Noran, Middleton, WI, USA). For LSCM an objective plan apochromat 63x/1.4 oil and for PCM an objective plan-neofluar 40x/0.75 (both Zeiss, Oberkochen, Germany) were used. The slit width for LSCM was 15 µm. Image processing for PCM occured with the setup described in [13] and the NIH image program [14] with the modification VETRAL [15]. All other computations were done with the computer program IGOR[16].

Fig. 1a shows LSCM micrographs taken parallel to horizontal sections through a DMPC vesicle containing polymerized actin. Fig. 1b exhibits the fluorescence intensity distributions along a line through the equatorial plane before (a) and after (b) polymerization. Clearly the actin is accumulated near the membrane after polymerisation forming a thin fluorescent shell of about 1 µm thickness. The formation of the actin cortex could also be monitored by characteristic shape changes of vesicles exhibiting excess bilayer area before polymerization. The originally flaccid and strongly flickering vesicles show buckling shape changes (similar to Fig. 2) within about 30 minutes after starting the polymerization.

Fig. 2a shows typical shape changes of a vesicle containing polymerized actin caused by thermal expansion of the bilayer. First the vesicle undergoes a buckling transition (cf. images a and b). This is reversible for small temperature increases of about 5 °C. During further heating a few or sometimes only one protrusions are formed which may consist of rounded blisters (cf. Fig 2a image c and d), cone-shaped protrusion (cf. Fig 2b image b), tubular protrusions (cf. Fig. 2b image d) and spherical buds connected with the mother vesicle by a neck (image not shown). The irreversible protrusions are stable for hours. The formation of only a small number of protrusions indicates that the excess area formed isotropically on the vesicle surface can slide over the cortex.

The protrusions may be partially stabilized by growing in of polymerized actin due to long time changes of the filament lengths.

About 10 % of the vesicles (such as that of Fig. 2) exhibit strong enough flickering in the quasi-spherical shape to be analysed by flicker spectroscopy [6]. Judged from previous studies [6] this is most likely due to the lateral tension generated by microbud formation. The flickering of the majority of the vesicles is, however, large enough to measure the mean square amplitudes $\sum \langle v_q^2 \rangle$ of the fluctuating membrane along the contour (cf. Fig. 3b). The bending modulus $k_c$ and the membrane tension $\sigma$ of the strongly flickering vesicles are determined by Fourier analysis of the fluctuation of the contour $v(\varphi,t)$ according to $v(\varphi,t) = R_0 \sum v_q(t) e^{-iq\varphi}$ ($\varphi$ azimuth angle, $R_0$ vesicle radius). The mean square Fourier amplitudes of the contour fluctuations (cf. Fig. 3) are related to the bending modulus $k_c$ and the normalized membrane tension $\sigma_{norm} = \sigma(\Delta) R_0^2 / k_c$ [17] according to [6, 17, 18]

$$\langle v_q^2 \rangle = \frac{k_B T}{k_c} \sum_{l=q}^{l_{max}} \frac{\left(P_l^q(\cos \pi/2)\right)^2}{(l+2)(l-1)[l(l+1) + \sigma_{norm}(\Delta)]} \quad , \quad (1)$$

where $P_l^q(\cos \pi/2)$ are the normalized associated Legendre polynoma. $l_{max}$ describes the highest order mode of membranes of thickness $d_m$ and is given by $l_{max} = 2\pi R d_m^{-1}$. Fitting of eq. (1) to the measured data of $\langle v_q^2 \rangle$ yields both the bending modulus $k_c$ and the membrane tension $\sigma(\Delta)$ [19]. An example is given in Fig. 3a. The interesting result is that $k_c$ is about equal to the value of that of the free bilayer containing the ionophore ($k_c = 5 \times 10^{-20}$ J) [6], which shows that the actin filaments are not tightly coupled to the bilayer. If the flickering is too weak to apply the rigorous analysis one can still estimate the mean square amplitude as a function of the azimuth angle of the contour (cf. Fig 3b).

The self-assembly of a thin shell of polymerized actin in a spherical vesicle can be understood in terms of the minimization of the bending energy of the semiflexible actin filaments. Actin polymerization in vitro leads to long filaments of average contour length $\langle L \rangle \cong 20$ μm [20] which is comparable to the radius of the vesicle studied. The average bending energy of a very thin shell of actin (with thickness $d_c \ll R$) composed of $N_F$ randomly oriented filaments is of the order

$$G_B \cong \frac{1}{2} N_F B \cdot \int_0^{2\pi R} R^{-2} ds \cong \pi N_F B R^{-1} \qquad (2)$$

where B is the filament bending modulus (dimension Jm). This equation is valid as long as the vesicle diameter is of simular size of or larger than the persistance length of the filaments, $L_p \cong 2$ μm [20b]. The total bending energy $G_B$ decreases as the radius of the shell increases, which is in contrast to the bending energy of the bilayer shell [21].

An important question is whether the flickering of the composite shell is determined by the undulations of the bilayer alone or of the whole shell. The following considerations favour the former possibility.

The effective bending modulus of a thin layer of $N_F$ randomly oriented filaments of length L is of the order $k_c^{eff} \cong N_F B L^{-1} / 2$. The number of filaments per vesicle of radius R can be expressed as

$$N_F = \frac{4\pi}{3} R^3 \cdot c_A N_a \cdot \frac{1}{\rho_A L} \tag{3}$$

where $c_A$ is the monomeric actin concentration, $\rho_A$ is the actin monomer number per length of a filament [22] and $N_a$ is Avogadro's number. The 2D effective bending modulus of the actin shell is then

$$k_c^{eff} \cong \frac{2\pi}{3} c_A N_A \cdot \frac{R^3}{\rho_A \cdot L^2} B \tag{4}.$$

For $L \cong R \cong 10 \mu m$ one finds (with $B \cong 4 \cdot 10^{-26}$ Jm [23]) $k_c^{eff} \cong 9 \cdot 10^{-18}$ J which is over an order of magnitude larger than the $k_c$-value measured (cf. fig. 3a), strongly suggesting that the actin cortex and the bilayer are decoupled. Further evidence for this comes from the finding that the bilayer can freely slide over the actin shell during the formation of blisters or buds after prolonged annealing. The distance $\langle h \rangle$ between the bilayer and the actin shell is below the optical resolution. $\langle h \rangle$ can be estimated by assuming that it is determined by the disjoining pressure by the bilayer undulation. In this case it is $\langle v^2 \rangle \approx \langle h \rangle^2$. The data of Fig. 3 yield a value of $\langle h \rangle \approx 0.5$ μm.

Another estimation of $\langle h \rangle$ is based on the assumption that the undulation-induced disjoining pressure, $P_{und}$, is balanced by the pressure $P_A$, exerted by the cortex towards the outside of the vesicle ($P_{und}=P_A$). The analysis of Fig. 3a shows that the disjoining pressure is dominated by the membrane tension since $\langle v_q^2 \rangle \propto q^{-1.3}$ instead of $\langle v_q^2 \rangle \propto q^{-3}$ characterisical for $\sigma = 0$ [6] Following Lipowsky [18] $P_{und}$ may then be approximately expressed as

$$P_{und} \approx 0.185 \cdot \frac{k_B T \, \sigma}{k_c} \cdot \exp\left(-\frac{h}{h_\sigma}\right) \cdot \left(\frac{h_\sigma}{h}\right)^{1/4} \left(4h^{-1} + h_\sigma^{-1}\right) \text{ with } h_\sigma \equiv \left(k_B T / 2\pi\sigma\right)^{1/2} \quad (5).$$

For a vesicle of R = 10 μm and $c_A$ = 2 μM (of fig. 2b images a and b) the pressure exerted by the actin cortex (with the single filament bending stiffness B = 4·10$^{-26}$ Jm) is of the order of $P_A$ = 4.5·10$^{-6}$ Jm$^{-3}$. It would be just balanced by the undulation pressure at a distance $\langle h \rangle = 0.4$ μm for $\sigma = 3\cdot 10^{-7}$ Jm$^{-2}$ (cf. Fig. 3a), in good agreement with the above value obtained by flicker analysis.

The self-assembly of a thin actin shell in giant vesicles exhibiting diameters comparable to the persistence length of the filaments is a consequence of the semiflexible nature of actin. Even if the actin is polymerized in large deflated vesicles they form quasi spherical shells while the excess bilayer area is transformed into buds which eventually detach (images not shown). We do not find tube like shapes unless the vesicles are small (R < 0.1 μm). Such shapes have been found for large vesicles if the polymerization was performed with very high actin concentrations and completely different polymerization conditions [24]. Recent studies by cryo electron microscopy (EM) and reconstruction of the actin-bilayer shells by EM-tomography [25] show that elongated vesicles are favoured with small radii R ≈ 0.1 μm.

The behaviour of the semiflexible actin in giant vesicles is strikingly different from that of microtubuli which form long protrusion if excess area is provided but which form also rings if the vesicle is under high tension [26].

The selfassembly of actin cortices in giant vesicles is a first step towards the design of more realistic models of cell plasma membranes. The composite shells exhibit shape changes such as buckling and (multiple) blister formation which are not found for pure lipid vesicles [27] but which are typical for cells. The buckling may be caused by the arrangement of actin into domains of parallel aligned filaments with the domain walls forming flexible hinges. This is indeed suggested by freeze fracture EM (unpublished results of authors' laboratory).

## Acknowledgements

The research was supported by the Deutsche Forschungsgemeinschaft (SFB266) and the National Science Foundation (Grant No. PHY8904035). Helpful discussions with A. Boulbitch, R. Bruinsma, R. Lipowsky, U. Seifert and M. Wortis are gratefully acknowledged.

# Figure Captions

**Fig. 1**

a) Fluorescence micrographs of spherical DMPC vesicle with reconstituted polymerized actin (initial concentration 7 µM) taken with LSCM. Horizontal sections (distance 3 µm) from the top to the equatorial plane are shown.

b) Plot of fluorescence intensity along the line AB shown on image f in figure 1a. The intensities before (**O**) and after polymerisation for 2 hours (**☐**) are given. The full width at half maximum of the observed maxima is about 1 µm.

**Fig. 2**

a) Phase contrast micrograph of buckling and blistering of composite shell (DMPC vesicle containing 7 µM of actin) during gradual increase of bilayer excess area by thermal expansion. (a) 26.0 °C, (b) 26.7 °C, (c) and (d) 32.8 °C).

b) Same for vesicles containing 2 µM (image a and b) showing cone shaped protrusion and 7 µM showing tube-like protrusion (images c and d).

**Fig. 3**

a) Normalized mean square Fourier amplitude $\langle v_q^2 \rangle$ plotted as a function of undulation wave number, q, of fluctuating contour of a vesicle with reconstituted actin network for two different actin concentrations: (a) 2 µM and (b) 7 µM. The radii of both vesicles are 10 µm and the temperature was 26.5 °C. Following Häckl et al. [18] the bending stiffnes, $k_c$, and the tension, $\sigma$, is obtained by fitting Eq (1) to the $\langle v_q^2 \rangle$-versus-q plots.

$$A: k_c = (7.5 \pm 1.9) \cdot 10^{-20} \, J, \; \sigma = (4.5 \pm 0.3) \cdot 10^{-7} \, Jm^{-2};$$
$$B: k_c = (4.0 \pm 1.1) \cdot 10^{-20} \, J, \; \sigma = (2.8 \pm 0.2) \cdot 10^{-7} \, Jm^{-2}$$

The excessively large amplitudes of modes q=5 and q=6 for case (a) could be caused by a slightly non spherical average vesicle shape.

b) Radial plot of root mean square amplitudes $\sqrt{\langle v^2 \rangle}$ of the flickering of the vesicle shown in Fig. 2b (image b).

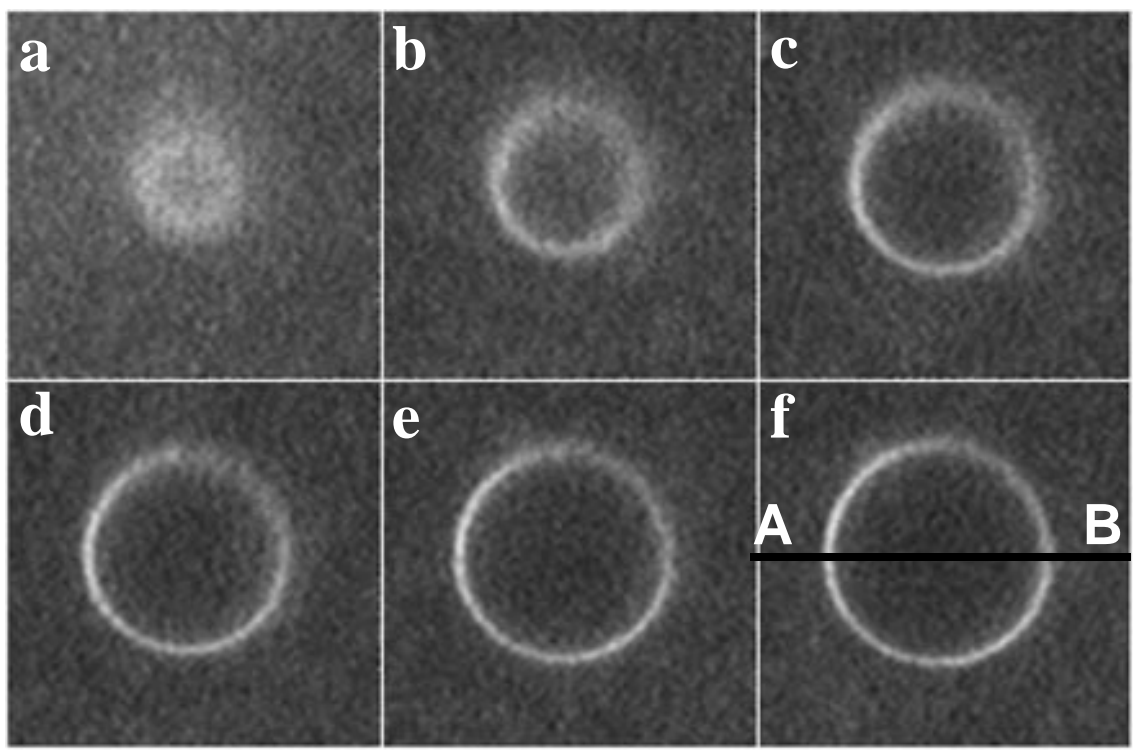

Fig. 1a
W. Häckl, M. Bärmann, E. Sackmann

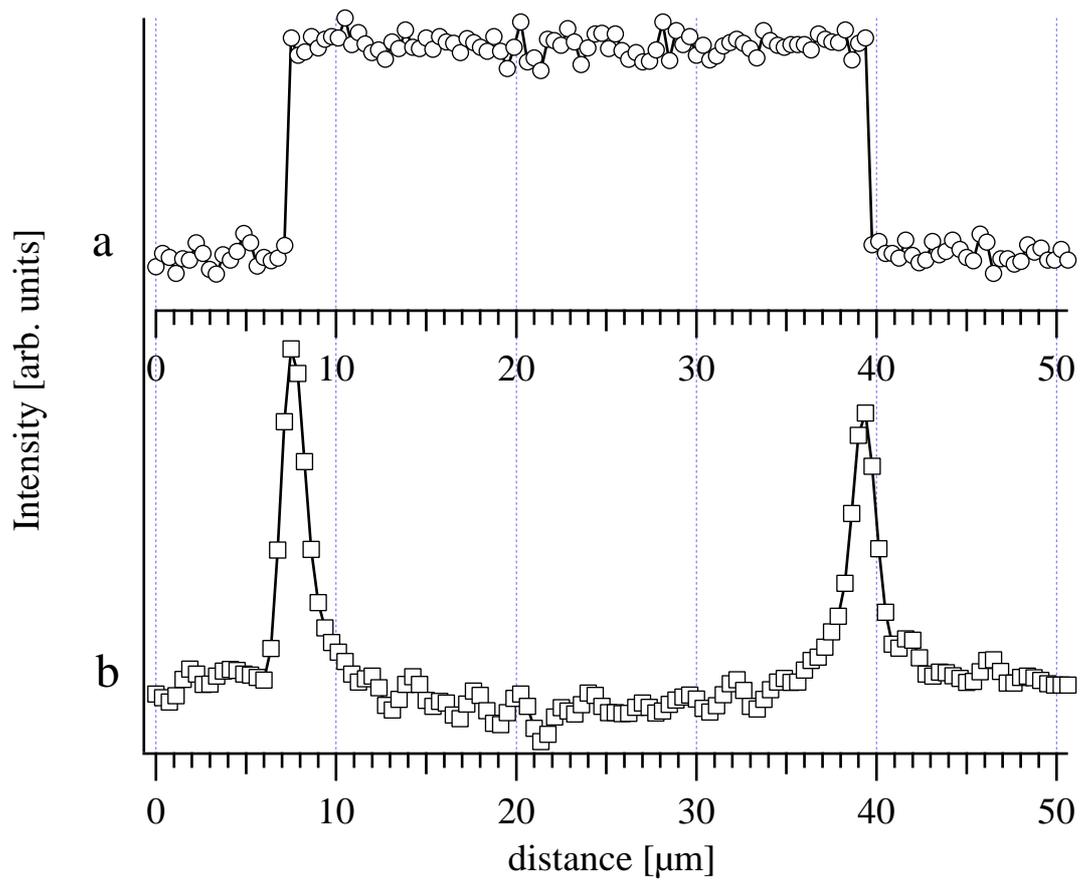

Fig. 1b
W. Häckl, M. Bärmann, E. Sackmann

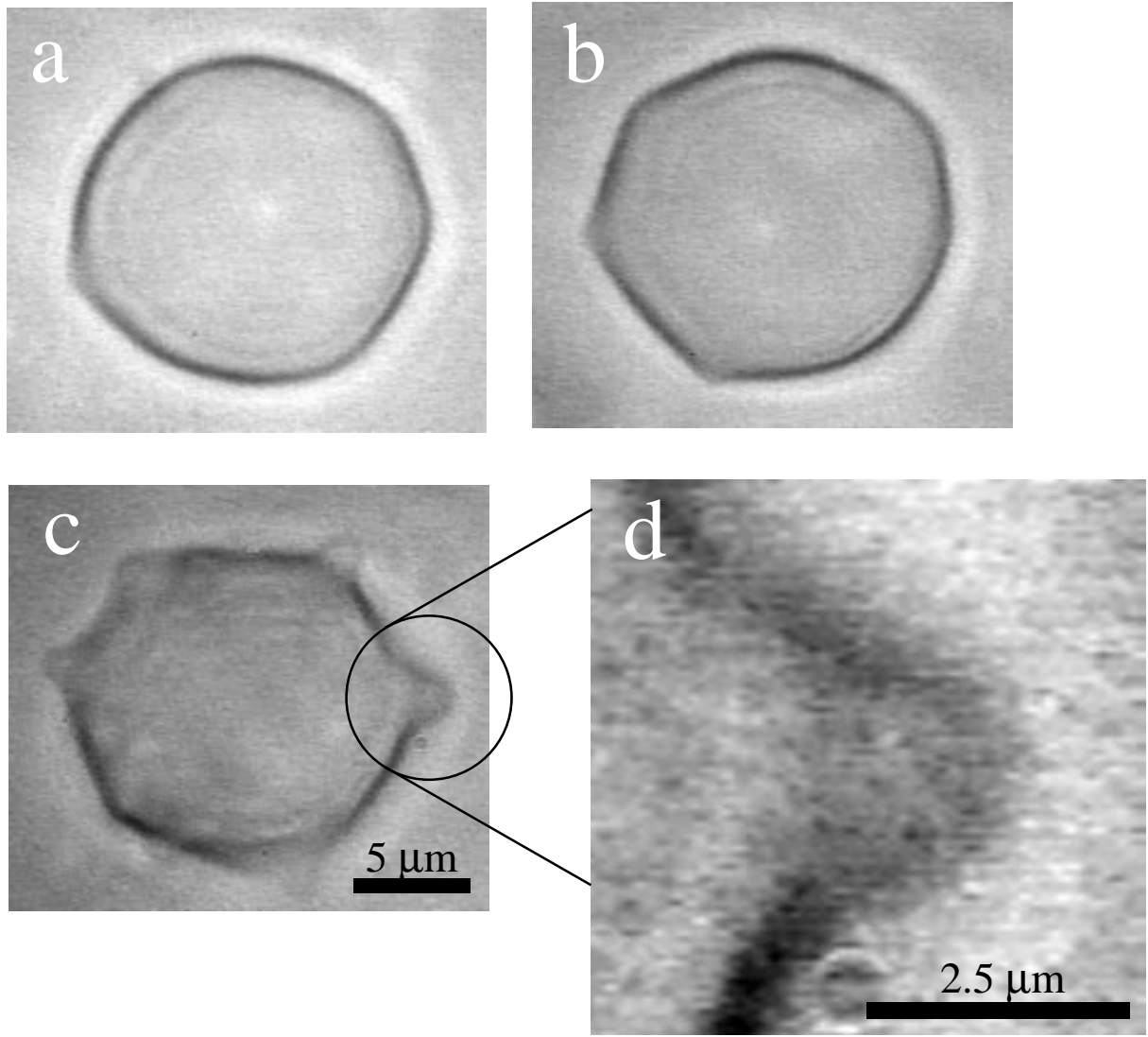

Fig. 2a
W. Häckl, M. Bärmann, E. Sackmann

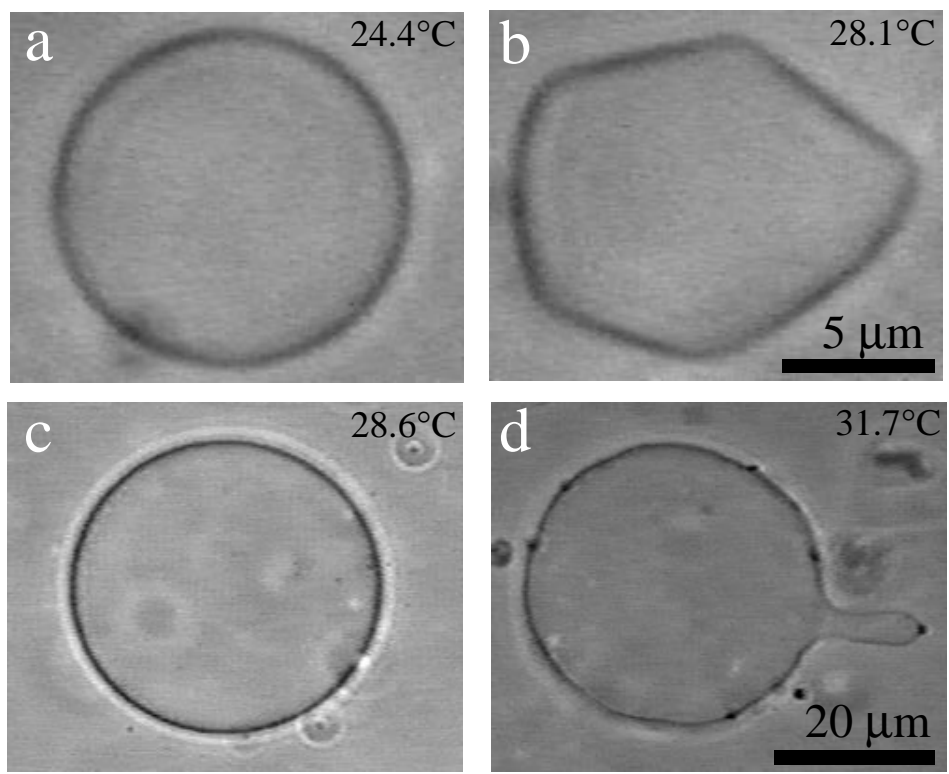

Fig. 2b
W. Häckl, M. Bärmann, E. Sackmann

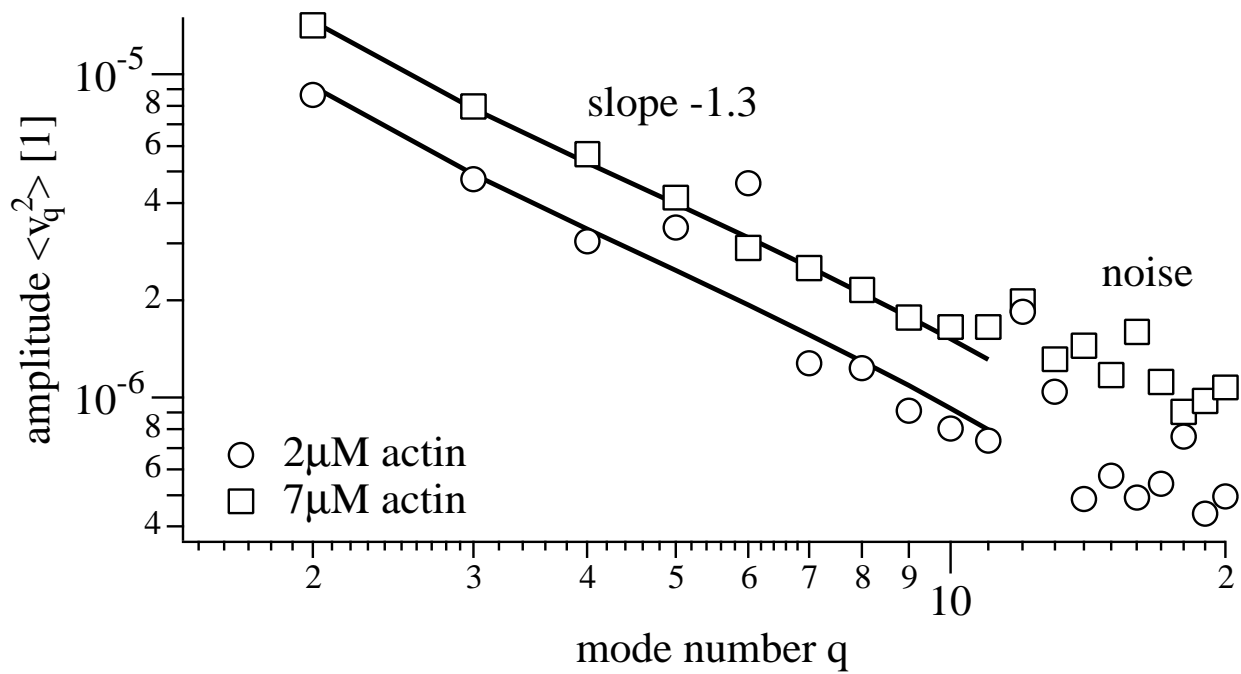

Fig. 3a
W. Häckl, M. Bärmann, E. Sackmann

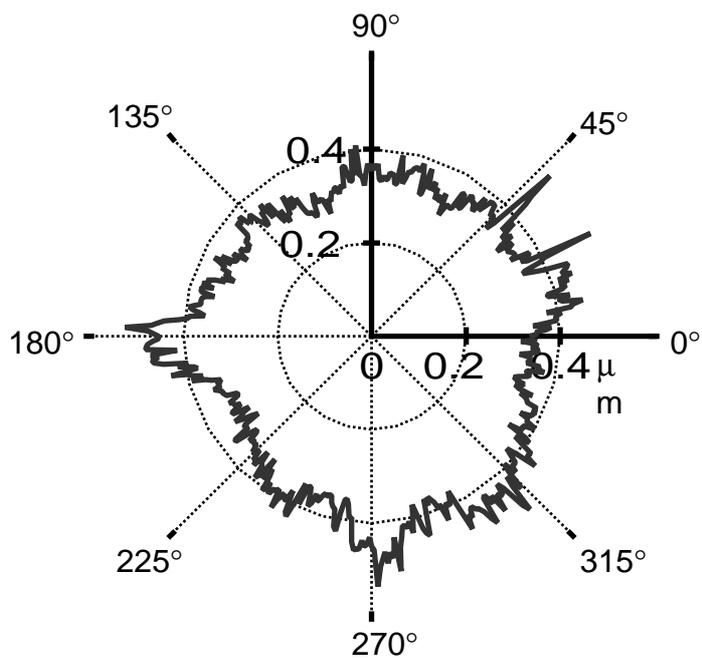

Fig. 3b
W. Häckl, M. Bärmann, E. Sackmann